\documentclass{aastex}
\usepackage{emulateapj5}

\begin{document}

\title{Galaxy Correlation Statistics of Mock Catalogs for
the DEEP2 Survey} \author{Alison L. Coil\altaffilmark{1}, Marc
Davis\altaffilmark{1}, Istvan Szapudi\altaffilmark{2}}
\altaffiltext{1}{Department of Astronomy, University of California,
Berkeley, CA 94720 -- 3411 \\ e-mail: {acoil@astro.berkeley.edu,
marc@astro.berkeley.edu}} \altaffiltext{2}{Institute for Astronomy,
University of Hawaii, 2680 Woodlawn Drive, Honolulu, HI 96822 \\
e-mail: {szapudi@IfA.hawaii.edu}}

\begin{abstract}

The DEEP2 project will obtain redshifts for $\sim$60,000 galaxies
between $z\simeq$ 0.7--1.5 in a comoving volume of roughly 7 10$^6$
Mpc$^3$ $h^{-3}$ for a $\Lambda$CDM universe.  The survey will map
four separate 2$^{\circ}$ by 0.5$^{\circ}$ strips of the sky. To study
the expected clustering within the survey volume, we have
constructed mock galaxy catalogs from the GIF and Hubble Volume
simulations developed by the Virgo consortium.  We present two- and
three-point correlation analyses of these mock galaxy catalogs to 
test how well we will measure these statistics, particularly in the 
presence of selection biases which will limit the surface density
of galaxies which we can select for spectroscopy.  We find that 
although the projected angular two-point correlation function $w(\theta)$
is strongly affected, neither the two-point nor three-point correlation
functions, $\xi(r)$ and $\zeta(r)$, are significantly 
compromised.  We will be able
to make simple corrections to account for the small amount of bias
introduced.  We quantify the expected redshift
distortions due to random orbital velocities of galaxies within groups
and clusters (``fingers of god'') on small scales of $\sim$1 Mpc
$h^{-1}$ using the pairwise velocity dispersion $\sigma_{12}$ and
galaxy-weighted velocity dispersion $\sigma_1$. We are able to measure
$\sigma_1$ to a precision of $\sim$10\%.
  We also estimate the
expected large-scale coherent infall of galaxies due to supercluster
formation (``Kaiser effect''), 
as determined by the quadrupole-to-monopole ratio
$\xi_2$/$\xi_0$ of $\xi(r_p,\pi)$.  From this measure we will be able
to constrain $\beta$ to within $\sim$0.1 at $z$=1.  

For the DEEP2 survey we will
combine the correlation statistics with galaxy observables such as
spectral type, morphology, absolute luminosity, and linewidth to
enable a measure of the relative biases in different galaxy
types.  Here we use a counts-in-cells analysis to measure $\sigma_8$
as a function of redshift and determine the relative bias between
galaxy samples based on absolute luminosity.  We expect to measure
$\sigma_8$ to within 10\% and detect the evolution of relative bias with 
redshift at the 4--5$\sigma$ level, with more precise measurements for the 
brighter galaxies in our survey.

\end{abstract}


\section{INTRODUCTION}

The DEEP2 survey \citep{dav00} will commence in the spring of 2002,
with the anticipated delivery of the DEIMOS spectrograph \citep{cow97}
to Keck in the fall of 2001.  The survey will use approximately 120
Keck nights over a period of three years to study galaxy properties,
evolution, and large-scale structure at $z\sim$ 1.  The DEEP2
survey will obtain spectra for $\sim$60,000 galaxies between
$z\approx$ 0.7--1.5.
The current plan is to use an 830 l mm$^{-1}$ grating with 0.75 $\arcsec$
slitwidths, resulting in a spectral resolution of 86 km s$^{-1}$.
We thus expect to easily measure linewidths to this limit and redshifts
to 10\% of this limit.  The survey will target four 2$^{\circ}$ by 
0.5$^{\circ}$ fields, which corresponds to a total comoving area of 
80 by 20 $h^{-2}$ Mpc$^2$ at
$z\sim$ 1 for a $\Lambda$CDM universe.  The line-of-sight comoving
distance for $z\approx$ 0.7--1.5 is $\sim$1400 $h^{-1}$ Mpc, though
in our magnitude-limited survey we will not sample the higher-redshift
region very densely.  The four fields were chosen as low-extinction
regions which are continuously observable from Hawaii over a six month
interval.  One field is the Groth Survey strip, and two fields are on
the equatorial strip which will be deeply imaged by the Sloan Digital
Sky Survey (SDSS).  Each of these fields are targets of a CFHT survey
conducted by Kaiser and Luppino, with the primary goal of deep imaging
in B, R, and I bands for weak lensing studies \citep{wil00}.  We will
use their photometric information to color-select galaxies with
$z\geq$ 0.7 and $m_I(AB)\leq$ 23.5 for the DEEP2 survey.  This
color-selection allows us to focus on the high-$z$ universe.

One of the key science goals of the DEEP2 survey is to measure the
two-point and higher-order correlation functions of galaxies at
$z\sim$ 1 as a function of other observables.  In almost all models
of structure formation (e.g., \citet{whi87}), galaxies are born as
highly biased tracers of the mass distribution, but their bias
diminishes with time. Spiral galaxies today appear to be weakly
biased, if at all, while the clustering of $z\sim$ 3 Lyman-break
galaxies requires a large bias for any reasonable cosmological model
\citep{gia98,wec98}.  Galaxies at $z\sim$ 1 should have an
intermediate degree of bias, with readily observable consequences.

The statistical clustering of galaxies at any epoch depends on both
the underlying dark matter distribution and the biasing function; the
two are degenerate without more information.  However, models suggest
that patterns exist in the {\it relative} biases between various types
of galaxies, and measuring these patterns as a function of absolute
luminosity, linewidth, or spectral type can break that degeneracy.
With sufficiently dense sampling, determining the clustering
properties of galaxies can yield direct measurements of their biasing.
For example, in linear biasing models [where $(\delta
\rho/\rho)_{gal}$ is assumed to be some constant $b \times (\delta
\rho/\rho)_{mat}$], the mean ratio of the skewness of the density
distribution function to the square of its variance has an expectation
value that scales as $1/b$ \citep{fry96,sco98}.  We will be able to
subdivide the DEEP2 sample as a function of internal linewidth and
estimate $b$ for each type of galaxy.  
The sampling density of the survey is such that we will
be able to obtain spectra for three-quarters of all galaxies down to
$L^*+0.5$ at the redshifts of interest.  Comparison of such measures
at $z\sim1$ to their values at $z=0$ will then be possible.

DEIMOS  has a field of view of 16$\arcmin$ by 5$\arcmin$ for both imaging
and multislit spectroscopy and will use an array of eight 2k by 4k
Lincoln Laboratory 30 $\micron$ thick CCDs, with an active
2-D flexure compensation system to minimize fringing at red
wavelengths.  As the DEEP2 survey will target four fields of 120$\arcmin$ by
30$\arcmin$, slitmasks will be made from a region of size 16$\arcmin$ 
by 4$\arcmin$ with two
rows of 60 masks each.  There will be $\sim$130 slitlets per mask,
with most of the slitlets aligned along the long axis, but with some
tilted as much as $30^\circ$ to track extended galaxies in order to
measure rotation curves.  
The mean surface density of candidate galaxies exceeds the number of
objects we can select, and spectra of selected targets cannot be
allowed to overlap on the CCD.  
Adjacent slitmasks will overlap by 2 $\arcmin$ so 
that each galaxy will have two chances to be selected for a mask.
Spectra will be obtained for $\sim$70$\%$ of
the targeted galaxies in the fields meeting our color and magnitude
selection criteria, chosen by a slitmask algorithm which is
necessarily biased against the highest-density regions, where the
spectra of nearby galaxies would overlap on the CCD.  Here we test
how this bias will affect our measurement of the underlying
 two-point and three-point correlation functions.

This paper uses a family of mock catalogs to model large-scale
structure studies that will be possible with the DEEP2 survey data.
Section 2 describes the construction of the mock catalogs while section
3.1 provides results of analysis of two-point correlation studies
in $\xi(r_p,\pi)$.  The high resolution spectroscopy of DEEP2 will allow
redshift space distortions to be readily measured, as discussed
 in section 3.2.  
This in turn will allow for a measurement of the small-scale thermal 
velocities $\sigma_{12}$ and $\sigma_1$, described in section 3.3.  
In section 3.4 we show that we can measure the three-point correlation
function $\zeta(r)$ with minimal bias.  Section 3.5 discusses the ability
to measure the evolution of clustering {\it within} the volume of the DEEP2
survey.  We expect to have the sensitivity to detect evolving bias and 
evolving $\sigma_8$.  Sections 4 and 5 present general discussion and 
conclusions.

\section{CONSTRUCTION OF MOCK CATALOGS}

We constructed mock catalogs from the Hubble Volume and GIF
simulations made by the Virgo Consortium
(www.MPA-Garching.MPG.DE/Virgo/).  
To match the DEEP2 survey, our mock catalogs cover 2$^{\circ}$ by
0.5$^{\circ}$ on the sky, with a redshift range of z=0.7--1.5.  
For both simulations we use the
$\Lambda$CDM models with $\Omega_{matter}$=0.3,
$\Omega_{\Lambda}$=0.7, $h$=0.7, and $\sigma_8$=0.9.  The Hubble
Volume Project produced deep-wedge geometry lightcone outputs from
simulations of 10$^9$ particles which include smooth clustering
evolution as a function of redshift \citep{evr00}.  The lightcone
output is designed to match what an observer would see and uses
propagating output filters at the speed of light through the
simulation in order to produce images which have evolution in the
redshift direction.  The $\Lambda$CDM simulation covered a comoving
cube of length 3000 $h^{-1}$ Mpc, and we constructed 16 independent
catalogs from subsets of the lightcone output.  The Hubble Volume
simulations have a mass resolution of 2.25 10$^{12}$ $h^{-1}$
$M_{solar}$ per particle and a softening scale of 100 kpc $h^{-1}$.
We only use particles in the simulation which are labeled
``galaxies'' in the L2 bias model, determined from the initial
overdensity field at that location as described in \citet{yos00}.
This bias model has both a lower and upper threshold cutoff in the
overdensity field inside of which ``galaxies'' will form.  This
results in an enhancement of voids as well as creates a small
anti-bias needed in $\Lambda$CDM models to match observed galaxy
correlations on small scales \citep{jen98}.

We also use simulations from the GIF project, which combine much
higher-resolution N-body simulations with semi-analytic models to
study formation and evolution of galaxies \cite{kau99}.  The particle
mass of the GIF simulations is 10$^{10}$ $M_{solar}$.  These
simulations are in the form of ``snapshots'' of comoving cubes at
various redshift intervals.  The cubes are 141 $h^{-1}$ Mpc on a side,
and we use outputs at z=0.62, 0.82, and 1.05.  In order to produce
catalogs which continuously cover z=0.7--1.5, we had to stack two
z=0.62 cubes, three z=0.82 cubes, and five z=1.05 cubes in the
redshift direction.  
Since the cubes can be randomly oriented, and 141
$h^{-1}$ Mpc corresponds to $\sim$ 5$^{\circ}$ at z=0.6 and
3.5$^{\circ}$ at z=1.0, we were able to select 2$^{\circ}$ by
0.5$^{\circ}$ subsets of different orientations from each cube when
stacking them, thereby reducing structure replication in the redshift
direction in our mock catalogs.  
We constructed 6 separate mock
catalogs from the GIF simulations, roughly equal to the number of
independent catalogs which could be made from the GIF galaxy
simulation output.  There is, however, some replication at $z \geq$
1.0, where we had to stack more cubes from the same snapshot output.

To assign each galaxy a redshift, we assumed a $\Lambda$CDM cosmology
with the following relation between comoving distance and redshift:
\begin{equation}
r=\frac{c}{H_0} \int \frac{dz}{\sqrt{(1+z)^3 \Omega_m + \Omega_\Lambda}}, 
\end{equation}
where $r$ is the comoving distance in units of $h^{-1}$ Mpc and
$\Omega_m$=0.3 and $\Omega_\Lambda$=0.7.  
To analyze both the Hubble Volume and GIF simulation mock catalogs
in redshift space we added the peculiar velocity along the line of sight,
$v_p$ (km s$^{-1}$), to the zero-peculiar-velocity-redshift, $z_0$: 
\begin{equation}
z = z_0+\frac{v_p}{c} (1+z_0).
\end{equation}
In the analysis of the redshift space catalogs, the relations between
$z$ and $r$ are inverted, and we convert the new $z$ to a given $r$.
Note that one must assume an underlying cosmology to analyze high-redshift
catalogs, and if the assumed model is incorrect the resulting statistics
will be distorted.

\begin{figure}
\epsscale{1.0}
\plotone{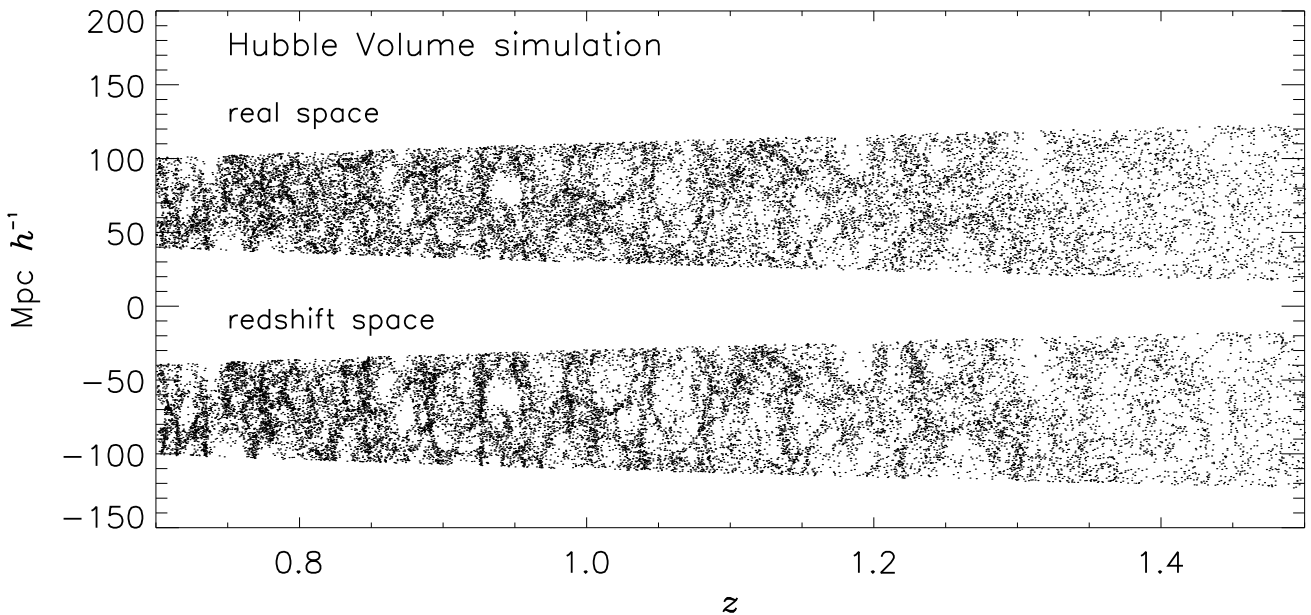}
\caption{One of our mock galaxy catalogs constructed
from the Virgo Consortium Hubble Volume simulations, 
shown in both real and redshift space. This lightcone output has
the advantage of continuous evolution, but the mass resolution is poor.
\label{hubblemock} }
\end{figure}

\begin{figure}
\epsscale{1.0}
\plottwo{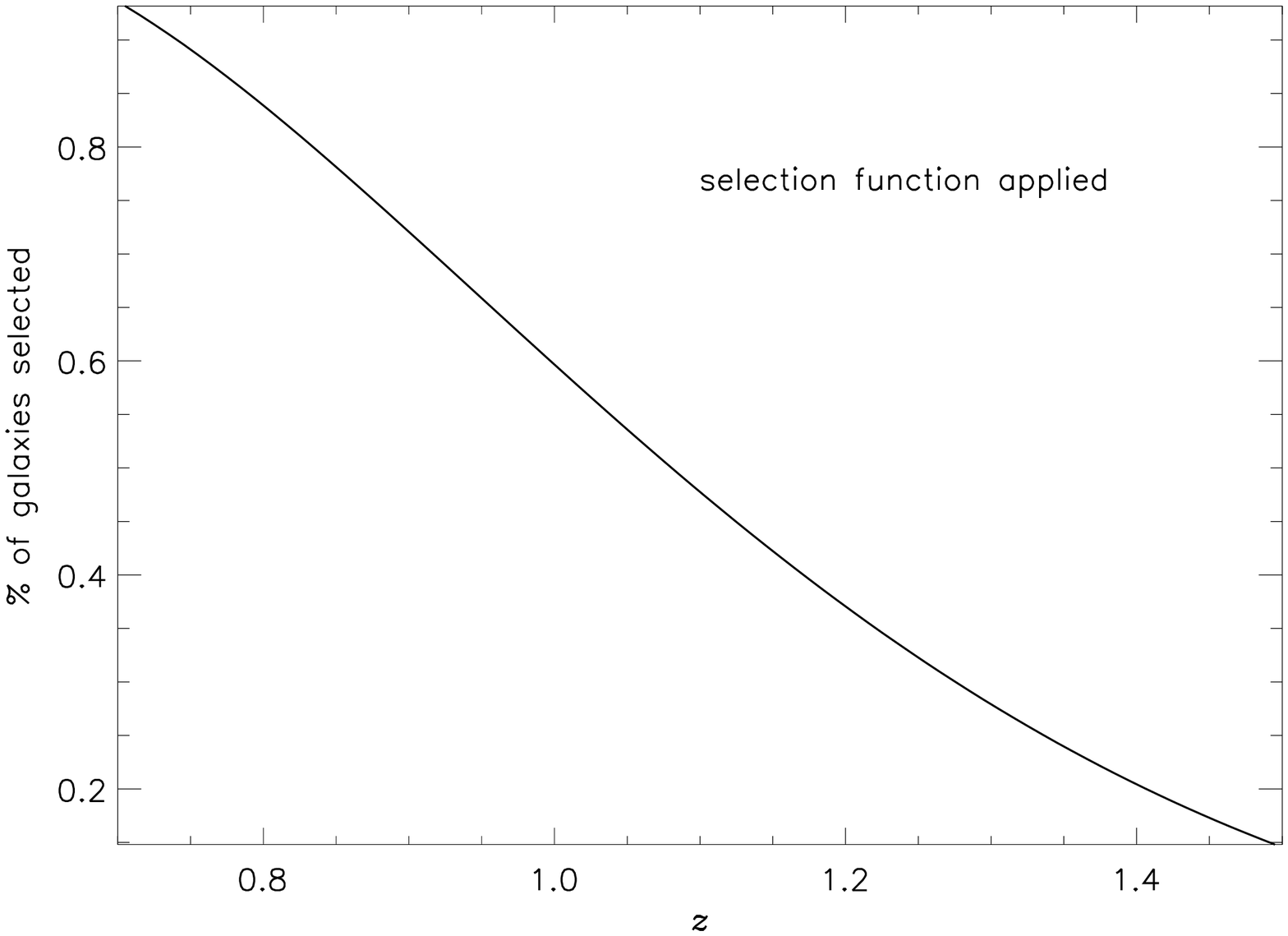}{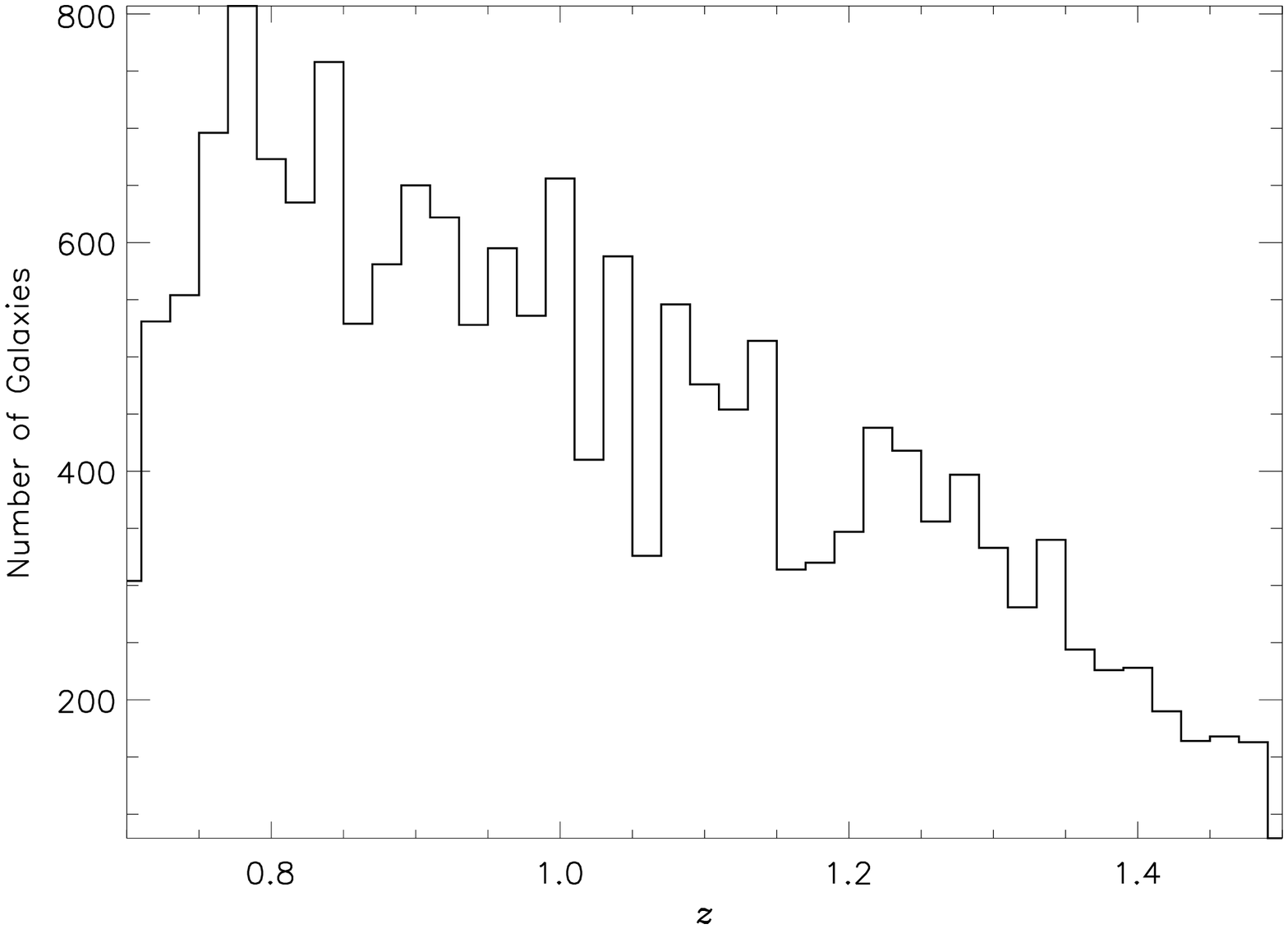}
\caption{$a$. The redshift 
selection function which was applied to the Hubble Volume mock catalogs.  
$b$. A histogram of the redshift distribution of galaxies
in one of the Hubble Volume  mock catalogs. \label{sf}}
\end{figure}

As the simulations provide a volume-limited sample, we apply a
selection function to our mock catalogs to mimic the $a$ $priori$
unknown selection function of the magnitude-limited DEEP2 survey.  For
the Hubble Volume simulations we follow the prescription of
\citet{pos98} for $N(z)$ models and integrate a Schechter luminosity
function in redshift bins.  We conservatively use a model with no
luminosity evolution as a pessimistic case from which to estimate the
correlation amplitudes and errors, as the no-evolution model has fewer
galaxies at high-$z$ than models with evolution.  After applying this
selection function, we then randomly dilute the density of galaxies
uniformly at all redshifts to match the observed surface density of
high-$z$ objects in our CFHT photometry fields of $\sim$ 5 galaxies
arcmin$^{-2}$.  We expect to successfully measure redshifts for
approximately 80$\%$ of our observed sample of galaxies in the survey,
and so we further dilute the mock catalog by an additional 20$\%$.
The resulting mock catalogs, each corresponding to one of our four fields, 
contain $\sim$18,000 galaxies. In Figure
\ref{hubblemock} we plot one of our mock Hubble Volume catalogs in
both real and redshift space, collapsed along the smallest axis.  The
no-evolution selection function applied to the mock catalogs is shown
in figure \ref{sf}$a$, along with the redshift distribution of
galaxies in one of the catalogs in figure \ref{sf}$b$.
%cellcounts_plotmock.pro creates figure 1
%selection_function.pro creates figure2a, also  figure2b.howto 

\begin{figure}
\epsscale{1.0}
\plotone{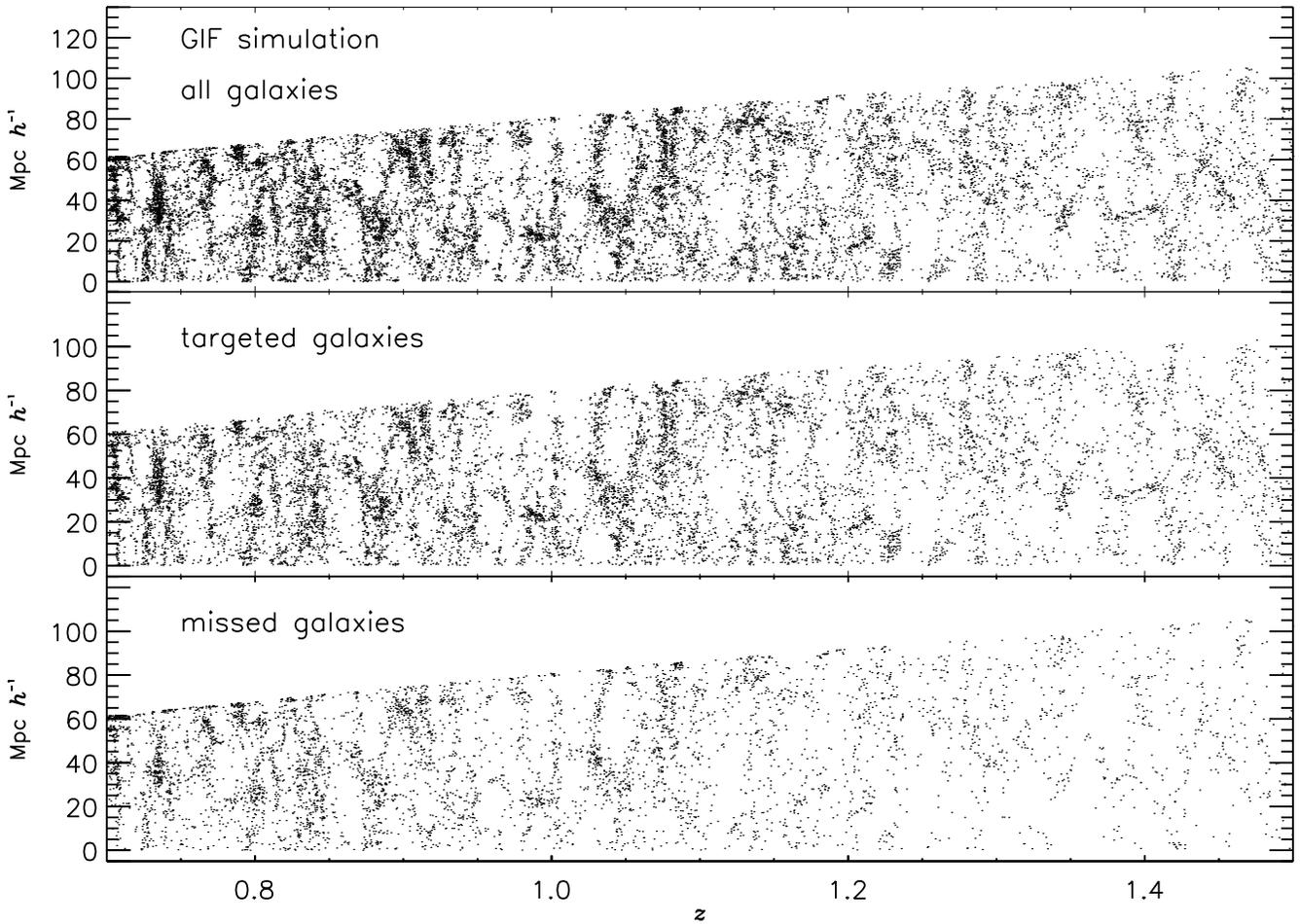}
\caption{One of our GIF mock galaxy catalogs shown in redshift space.  
In the top panel are all the galaxies in this mock catalog, while only
those targeted for slitmasks are shown in the middle panel.  The galaxies
for which we would not obtain spectra are in the lower panel.
This simulation has better mass resolution than the Hubble Volume simulation
and has semi-analytic models applied.  We used the known absolute 
luminosity of each galaxy to create a flux-limited sample of $I$=23.4 mag. 
  \label{gifmock}}
\end{figure}

The GIF simulations contain absolute magnitudes in $B$, $V,$ $R$, $I$
and $K$ for each galaxy, which we use to create a flux-limited
subsample from our mock catalogs.  We make an apparent $B$-band cut of
23.4 mag, which at $z \sim$1 corresponds to an apparent $I$-band
mag with the appropriate K-correction, resulting in flux-limited
catalogs of $\sim$19,000 galaxies.  An image of one of our GIF mock
catalogs is shown in figure \ref{gifmock} in redshift
space.  The redshift distribution of these catalogs agrees quite well
with that of the Hubble Volume mock catalogs.  Figure \ref{gifmock}
also shows the results of our target selection on one of the GIF mock
galaxy catalogs.  The middle panel shows galaxies which would be targeted
to be on a slitmask, while the lower panel shows galaxies for which
 we would not
be able to obtain spectra.  The map of galaxies missed by our selection
procedure traces the same structure as the map of our targeted galaxies.

\section{RESULTS}

The GIF simulations match our current CFHT photometry quite well in
terms of the projected angular correlation $w(\theta)$, notably at
small separations of $\leq$20 arcsec.  This holds for all the galaxies
as well as subsamples selected by color.  We therefore use the GIF
mock catalogs to estimate how well we will measure the two- and
three-point correlation functions and their redshift distortions, and to
test the effects of our target selection algorithm, which is most
strongly biased on small projected scales on the sky.  The Hubble
Volume simulations do not have the mass resolution to match the
projected correlation amplitudes on small scales but do have the
advantage of continuous clustering evolution with redshift.  We use
both the Hubble Volume and GIF simulations to estimate how well we
will measure $\sigma_8$ as a function of redshift in the DEEP2 survey.

\subsection{The Two-point Correlation Function and Bias of our Slitmask Algorithm}

\begin{figure}
\epsscale{0.7}
\plotone{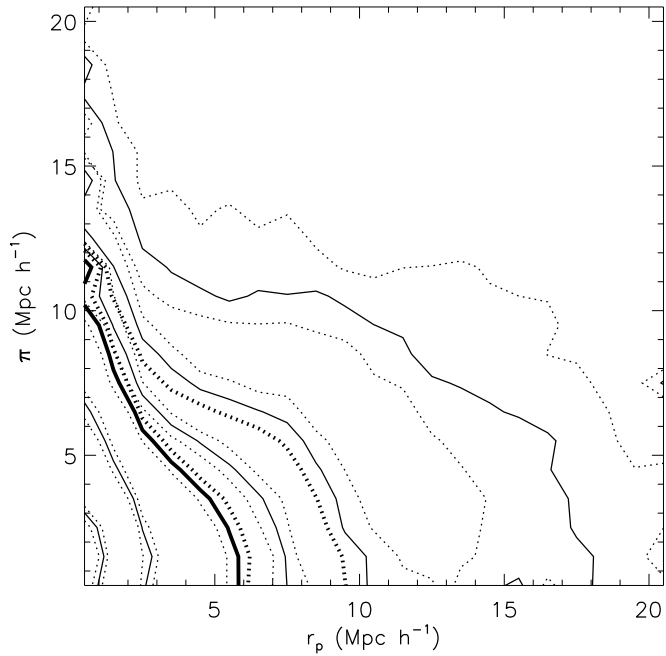}
\caption{The mean redshift-space two-point correlation function $\xi(r_p,\pi)$ of 
our six GIF mock galaxy catalogs is shown in solid lines, with 1$\sigma$ 
errors in dotted lines. The data have been smoothed with a 3-pixel boxcar
in each direction. 
The contour levels are 0.25, 0.5, 0.75, 1.0 
(bold contour), 2.0, and 5.0. The scale length of the clustering is 
$\sim$ 6 $h^{-1}$ Mpc, as seen on the x-axis.  \label{xisp.mean}}
\end{figure}

The two-point correlation function is defined as the excess
probability above random that a pair of galaxies exists with a given
separation (see Peebles 1980 for details).  This statistic is a simple
way to characterize the amount of clustering seen in the galaxy
distribution.  We present in figure \ref{xisp.mean} the average
two-point correlation analysis $\xi(r_p,\pi)$ of the six GIF mock
galaxy catalogs, as a function of separation across the line of sight
($r_p$) and along the line of sight ($\pi$).  The thick solid contour
traces where the correlation amplitude is equal to 1.0, and the scale
length of the clustering is roughly 6.0 $h^{-1}$ Mpc in real space as
seen on the $r_p$-axis.  The dotted contours are 1$\sigma$ standard
deviations derived from the six independent catalogs.  These contours
are quite confined, confirming that the DEEP2 design will lead to
strong constraints on the clustering of distant galaxies.  There is,
however, significant covariance among the residuals.  The covariance
matrix is comprised of correlation coefficients $r_{ij}$
\begin{equation}
r_{ij} \equiv \frac{\sigma_{ij}}{\sigma_i \sigma_j}
\end{equation}
where $\sigma_{ij}$ is the covariance between points $i$ and $j$ and
$\sigma_i$ is the square root of the variance of point $i$.  The
values of the covariance matrix are high, above 0.75 for scales of 5
-- 20 $h^{-1}$ Mpc.  This is not surprising, suggesting that poorly
sampled large-scale modes dominate the errors.  %The expected
%covariance of the DEEP2 data on large scales may be slightly lower, as
%the data will be taken from regions on the sky which are more widely
%separated than the mock catalogs.  
Error analysis in the real data will need to account for this covariance.
%The high covariance suggests that
%we may need to bin the data until the data points are independent.

\begin{figure}
\epsscale{0.6}
\plotone{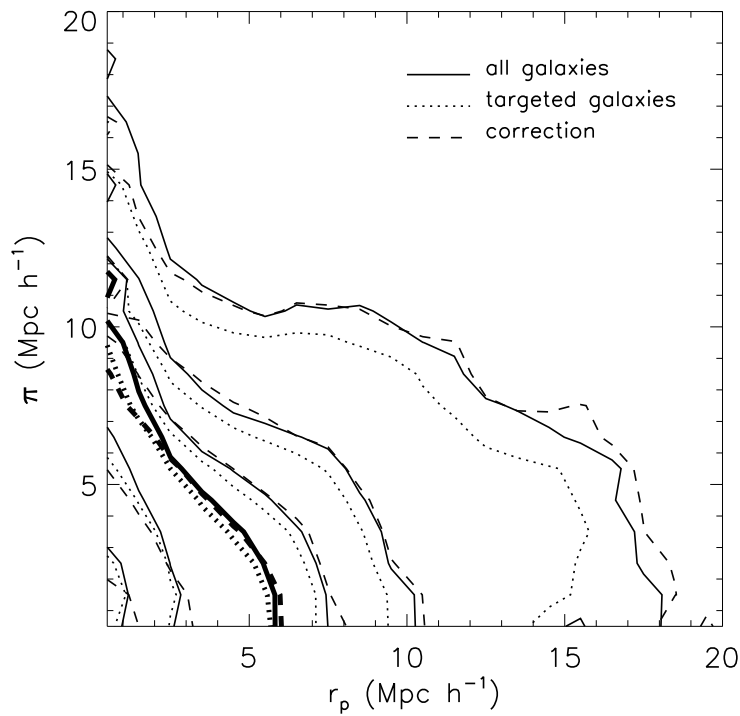}
\caption{The effects of our target selection algorithm on the two-point 
correlation amplitude on our GIF mock catalogs.   The 
solid contours show $\xi(r_p,\pi)$ for all the galaxies in the catalog,
with contours levels of 0.25, 0.5, 0.75, 1.0 (bold), 2.0, 5.0.  
The data have been smoothed with a 3-pixel boxcar in each direction. 
The dotted
contours are for only those galaxies selected to be observed. 
$\xi(r_p,\pi)$ of the targeted galaxies underestimates 
the true correlation amplitude.
Dashed contours show a simple correction for untargeted galaxies where the 
redshift of the nearest neighbor on the sky within the expected 
photometric redshift error is used for the untargeted galaxy.
   \label{xisp.mask}}
\end{figure}

A critical function of the GIF mock catalogs is to test the effects of the
target-selection algorithm on the measured two-point correlation function.
The results are shown in Figure \ref{xisp.mask}.  Here 
we plot the mean two-point correlation
amplitude $\xi(r_p,\pi)$ for all the galaxies in our catalogs in solid
contours as well as for only those galaxies selected to be on
slitmasks in dotted contours.  The correlation amplitude for targeted
galaxies alone is lower than for all the galaxies, as expected.  We
also show in figure \ref{xisp.mask} the result of a simple correction
for this bias.  For galaxies not targeted to be on slitmasks,
we substitute the redshift of the nearest neighbor on the sky which was
targeted and is within the expected photometric redshift error of $z$=0.1 
of the un-targeted galaxy and use its redshift as that of the 
un-targeted galaxy.  In
this way we use the angular information and the photometric redshift  
of the un-targeted galaxies and can partially recover the amplitude 
errors in the correlation analysis.  

This correction works extremely well on large scales, $\ge$5 $h^{-1}$ Mpc. 
The correction underestimates the ``finger of god''
effect on small scales ($\leq$5 $h^{-1}$ Mpc), where it effectively positions
the un-targeted galaxies at the identical physical distances of the 
 nearest neighbors on the sky, thereby reducing the effect of 
random thermal motion. The conclusion is that
the target-selection bias can be corrected in a simple manner
on scales $\ge$5 $h^{-1}$ Mpc.  The most
straightforward means of dealing with the resulting bias on small
scales will probably
be to filter all models by the same selection procedure, to estimate
the degree to which the targeted galaxies are underestimating the
correlation amplitude.

\subsection{Redshift Distortions in the Measured Two-point Correlation Function}

The mock catalogs can be used to quantify the precision with which we
can  measure
redshift distortions in the two-point correlation function $\xi(r_p,\pi)$
in the DEEP2 survey.  Redshift distortions are expected on small
scales due to the random internal velocities of galaxies in groups and
clusters, creating so-called ``fingers of god'' on small scales
($\leq$ 5 $h^{-1}$ Mpc).  This effect can be seen in the redshift
two-point correlation function $\xi(r_p,\pi)$ shown in figure
\ref{xisp.mean} as the elongation along the y-axis (separation along
the line of sight) at small scales.  On larger scales a flattening of
the two-point correlation function contours can be seen, which is due to
coherent infall of galaxies resulting from the gravitational pull of
large forming superclusters.  A standard method by which to quantify
these redshift-space distortions is to measure $\xi_2$/$\xi_0$, the
quadrupole to monopole moments of the two-point correlation function
\citep{ham98}.  The multipole moments of the two-point correlation
function are defined as
\begin{equation}
\xi_l(s) = (2l+1)/2 \int{\xi(r_p,\pi) P_l(\cos\theta) d\cos\theta}.
\end{equation}

\begin{figure}
\epsscale{0.8}
\plotone{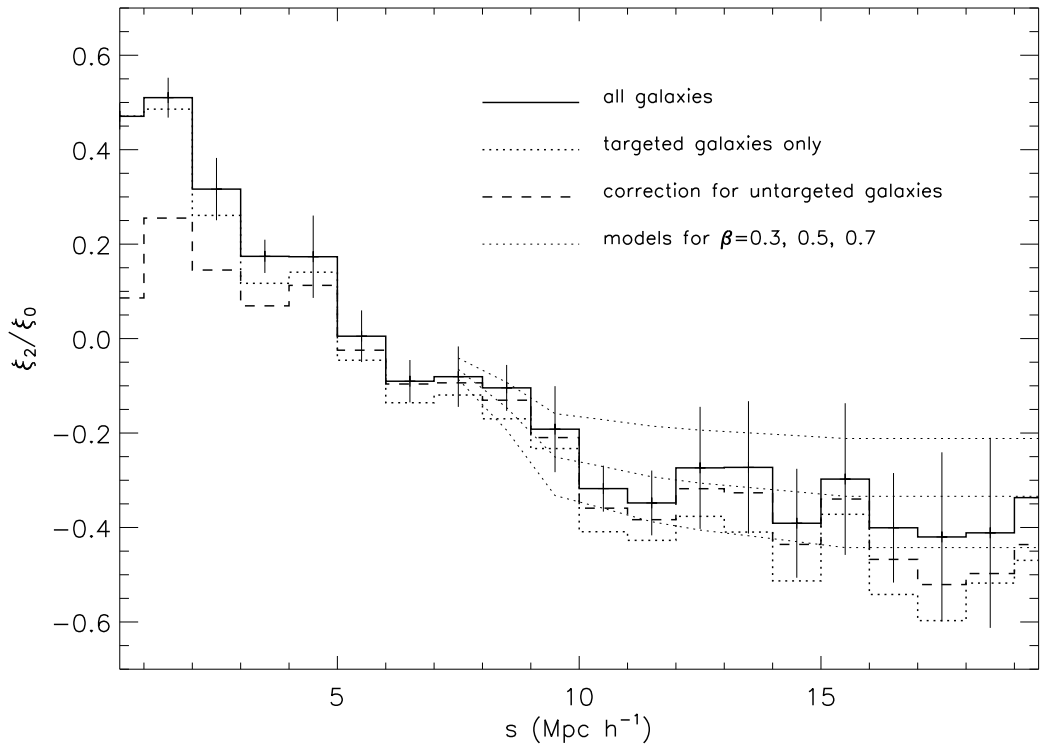}
\caption{Redshift distortions seen in the two-point correlation function 
are quantified by the quadrupole-to-monopole ratio $\xi_2$/$\xi_0$ of 
$\xi(r_p,\pi)$, plotted for all the galaxies in the mock catalogs
as a solid line with 1$\sigma$ errors.    
The dash-dot line is for only those galaxies selected 
to be observed on slitmasks, and the dashed line show the results of a simple 
correction for untargeted galaxies.  Dotted lines show models of
$\xi_2$/$\xi_0$ for different values of $\beta$. \label{gifQ}}
\end{figure}

$\xi_2$/$\xi_0$ is greater than zero on small scales where the
``fingers of god'' are apparent and less than zero on large scales
where coherent infall appears.  Figure \ref{gifQ} plots
$\xi_2$/$\xi_0$ as a function of redshift separation, with 1$\sigma$
errors as derived from the six separate GIF mock catalogs.  Unlike the
measurement of $\xi(r_p,\pi)$, the covariance matrix is large only on large
scales, greater than 12 $h^{-1}$ Mpc.  The bias of our target
selection procedure can be seen as the dotted line, and the correction
discussed in the last section is shown with a dashed line.  As with
the $\xi(r_p,\pi)$ statistic, the effect of target selection is to
slightly decease the amplitude of $\xi_2$/$\xi_0$ on all scales. 
The applied correction increases $\xi_2$/$\xi_0$ on scales
$\ge$5 h$^{-1}$ Mpc and underestimates it on small scales. We will need to
apply a bias correction to account for this in the survey.  However, the mock
catalogs show clearly that the correlation anisotropy will be robustly
detected in the DEEP2 survey.

The amount of flattening seen on large scales constrains the parameter
$\beta \equiv \frac{\Omega_m^{0.6}}{b}$, where $b$ is the
 bias between the galaxy and dark matter clustering.  In linear
theory, 
\begin{equation}
\xi_2/\xi_0=f(n) \frac{\frac{4}{3}\beta+\frac{4}{7}\beta^2}{1+\frac{2}{3}\beta+\frac{1}{5}\beta^2}
\end{equation}
where $f(n)$ =(3+$n)/n$ and $n$ is the index of the two-point 
correlation function in a power-law form: $\xi\propto r^{-(3+n)}$ 
\citep{ham92}.  Our 1$\sigma$ errors
plotted in figure \ref{gifQ} show that in one of our slices we
should be able to measure $\beta$ to within +/-- 0.2, so that for our
full sample of four slices our error on $\beta$ should be +/-- 0.1.
In order to extract $\Omega_m$ from our measurement of $\beta$ we will
need to know the {\it absolute} bias of our galaxy sample, which 
we may be able to determine from our measurements of the three-point 
correlation function relative to the two-point function (see section 3.4).

\subsection{Pairwise and Object-weighted Velocity Dispersion}

The pairwise velocity dispersion is a parameter which measures the
small-scale thermal motions of galaxies and probes the mass
density of the universe.  It has been measured to be around 400-500 km
s$^{-1}$ for various redshift surveys at low-$z$ (e.g. Fisher et al. 1994; 
Marzke et al. 1995; Jing, Mo, \& B\"{o}rner 1998). 
\nocite{jin98,fis94,mar95}  In this analysis, we confine our
test to a measurement of the pairwise velocity dispersion
$\sigma_{12}$ %in the GIF mock galaxy catalogs 
by comparing
$\xi(r_p,\pi)$ in real and redshift space on the scale of $r_p$=1
$h^{-1}$ Mpc, where the ``finger of god'' effect dominates redshift
distortions.  (In this section we will use a subscript R to indicate
real space.)  For this analysis we use the flux-limited mock catalog before
slit selection, averaged over all 6 catalogs.
  
 Following \citet{fis94}, we first define
\begin{equation}
\xi(r_p=1,\pi) \equiv 0.5[\xi(r_p=0.5\: \mathrm{Mpc}\,
h^{-1},\pi)+\xi(r_p=1.5 \:\mathrm{Mpc} \,h^{-1},\pi)]
\end{equation}
in both real and redshift space, $\xi_R(r_p=1,\pi)$ and
$\xi(r_p=1,\pi)$, for values of $\pi \le$ 20 $h^{-1}$ Mpc.  
We then normalize $\xi(r_p=1,\pi)$ so that our
subsequent fitting will be sensitive to the overall shape of
$\xi(r_p=1,\pi)$ but insensitive to the amplitude:
\begin{equation}
\xi(\pi)=\frac{\xi(r_p=1,\pi)}{\int_{0}^{\pi_{max}}\xi(r_p=1,\pi)
\,d\pi}
\end{equation}
where $\pi_{max}$ is the maximum value of $\pi$ used in the analysis
(here $\pi_{max}$=20 $h^{-1}$ Mpc).

\begin{figure}
\epsscale{0.8}
\plotone{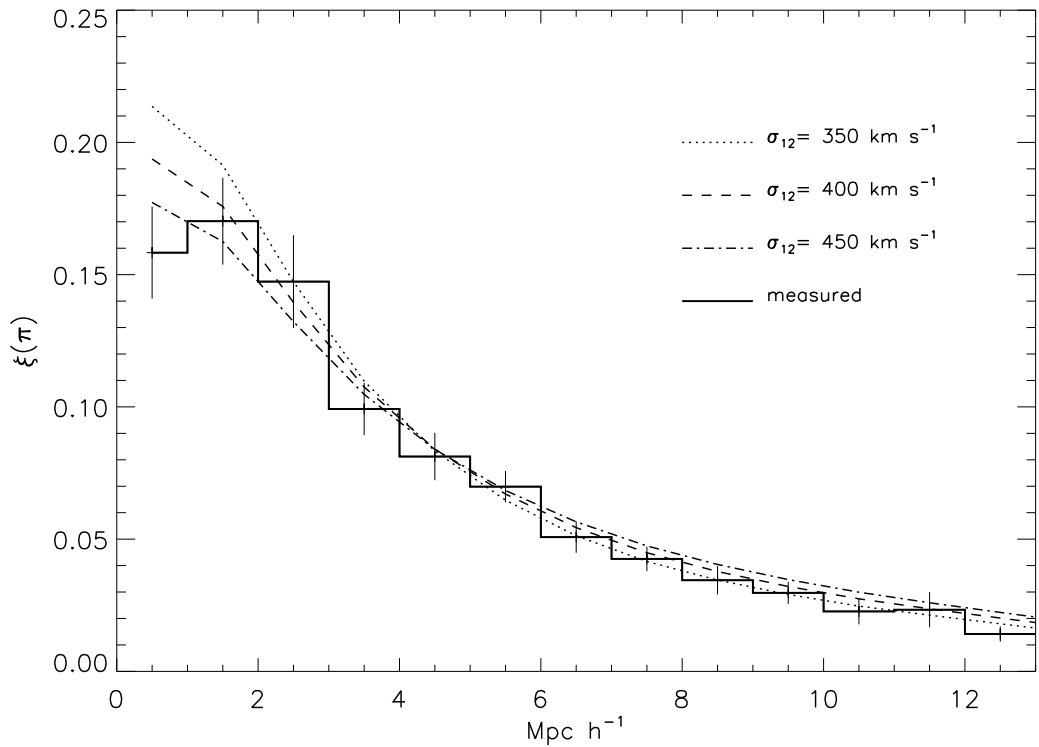}
\caption{Measured $\xi(\pi)$ in redshift space compared to models with
different pairwise velocity dispersions. The best fit is 
$\sigma_{12}$=410 +/--80 km s$^{-1}$.  \label{sigma12}}
\end{figure}

Within the mock catalogs, the pair-weighted dispersion of the
known peculiar velocities is 500 km s$^{-1}$ for
galaxies with radial separations $\le$20 $h^{-1}$ Mpc within a projected
distance of 1 $h^{-1}$ Mpc, as measured
in real space.  The mean infall velocity,
$v_{12}$, is 160 km s$^{-1}$ at the same scale.
To determine $\sigma_{12}$ in the redshift space catalog, 
we construct models of $\xi(\pi)$ by
convolving the measured $\xi_R(\pi)$ in real space with different
distribution functions of velocity differences for pairs of galaxies
separated by vector distance $\mathbf{r}$ to compare with the measured
$\xi(\pi)$.  
Let the vector $\mathbf{r}$ separating two galaxies be
decomposed into a transverse component $r_p$ and a line-of-sight
component $y$, where $y$ can be converted into a velocity separation
at the given redshift.  Then the velocity difference is given as 
$\pi$-$y$, and the convolution becomes
\begin{equation}
1+\xi(\pi)=\frac{1}{\sqrt{2}}\int \frac{dy}{\sigma_{12}}\:
(1+\xi_R(\pi)) \:exp(-\sqrt{2}\left|\frac{\pi-y-yv_{12}(r)/r}{\sigma_{12}}\right|),
\end{equation}
where $r^2=r_p^2+y^2$. For the mean infall velocity, $v_{12}(r)$, we
use the similiarity solution of \cite{dav83}, $v_{12} \sim r/[1+(r/r_0)^2]$ 
with $r_0$=5.83 h$^{-1}$ Mpc. At $z$=1 in a LCDM model this corresponds
to 71 km s$^{-1}$.
Figure \ref{sigma12} plots the measured $\xi(\pi)$ with models of
$\sigma_{12}$ equal to 350, 400, and 450 km s$^{-1}$.  Using a
minimum chi-squared test, we find that the best fit is  $\sigma_{12}$=410 km
s$^{-1}$ with a 1$\sigma$ error of 80 km s$^{-1}$. 
If we instead use a mean infall velocity of $v_{12}(r)$=160 km s$^{-1}$
at $r$=1 $h^{-1}$ Mpc, we find $\sigma_{12}$=485 km s$^{-1}$, consistent 
with the measured peculiar velocity
dispersion in the mock catalog of 492 km/s.  Note that redshift space
distortions  are larger at high redshift than at low redshift by the factor
$(1+z)$, and that we have removed this extra factor in our modeling of 
 $\sigma_{12}$ by simply using 1+$z$=2.   

\begin{figure}
\epsscale{0.7}
\plotone{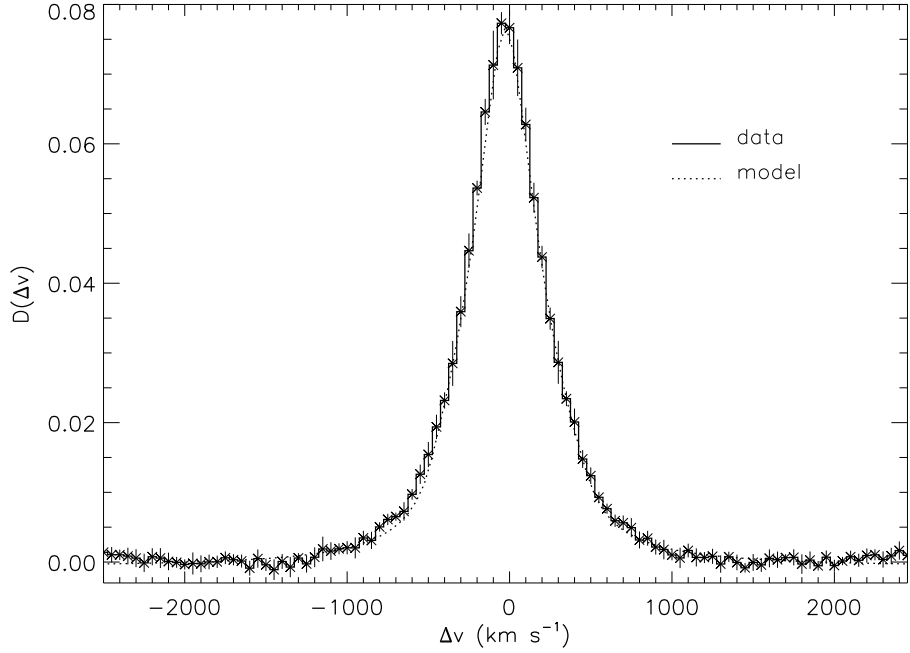}
\caption{Galaxy-weighted velocity distribution $D(\Delta v)$ 
for the six GIF mock catalogs. The best fit model shown has 
$\sigma_1$=180 km s$^{-1}$.
 \label{sigma1}}
\end{figure}

The $\sigma_{12}$ statistic is known to be unstable, however, as it is
pair-weighted and therefore quite sensitive to rare, rich galaxy
clusters \citep{dav97}.  \citet{jin982} also argue that the measured
value of $\sigma_{12}$ depends strongly on the assumed value of
$v_{12}$, the mean infall velocity, which here we have modeled in
an adhoc manner. We therefore also
measure the galaxy-weighted velocity dispersion $\sigma_1$, which is a
more stable statistic and a measure of the one-dimensional peculiar
velocity dispersion of galaxies relative to their neighbors.  This
statistic is more easily evaluated and does not rely on assumed values
of the mean infall.  Following the prescription of Baker, Davis, and
Lin (2000) \nocite{bak00} we construct the velocity distribution $D
(\Delta v)$ by a weighted sum over $N_g$ galaxies:
\begin{equation}
D(\Delta v)=\frac{1}{N_g}\sum_{i=1}^{N_g} w_i\left[P_i(\Delta
v)-B_i(\Delta v) \right]
\end{equation}
where $w_i$ is the weight for each galaxy, determined by the number of
neighbors in excess of a random background.  Here $P_i(\Delta v)$ is
the distribution of neighboring galaxies within a cylinder of
projected radius $r_p$ and half-length $v_l$ in redshift space, while
$B_i(\Delta v)$ is the background distribution expected for an
unclustered galaxy distribution.  \citet{bak00} describe a weighting
scheme where galaxies for which the distribution is negative can be
incorporated into $D (\Delta v)$ by considering the high and low
density samples separately and combining them, weighting by the number
of objects included in each sample.  This statistic results in an
unbiased, object-weighted measure of the thermal energy of the galaxy
distribution.  The velocity distribution $D (\Delta v)$ can be seen in
figure \ref{sigma1}.  To measure the intrinsic dispersion, $\sigma_1$,
we model the real space correlation function $\xi(r)$ convolved with
an exponential velocity broadening function,
\begin{equation}
f(v)=\frac{1}{\sigma_1}exp(-\frac{|v|}{\sigma_1})
\end{equation}
(see \cite{dav97} and \cite{bak00} for details).  We measure
$\sigma_1$=180 +/--20 km s$^{-1}$ for a $r_p=1$ $h^{-1}$ Mpc and
$v_l=$2500 km s$^{-1}$.  We find similar results for $r_p=$2 $h^{-1}$
Mpc.  The model fit for $\sigma_1$=180 km s$^{-1}$ is shown in figure
\ref{sigma1}.  

This value of $\sigma_1$ should be compared to the
1-D rms peculiar motion of 315 km s$^{-1}$ 
for the GIF mock galaxies in the catalog.
This rms peculiar motion includes large scale flows, while $\sigma_1$ 
 measures only the nonlinear thermal component of the peculiar velocities.
The difference is of the magnitude expected.  Note that \cite{bak00}  
measured $\sigma_1=126 \pm 10$ km s$^{-1}$ for the LCRS survey, and it is 
quite likely that DEEP2 will report a value lower than this for the distant
Universe.  This might become a modest mismatch between 
the GIF simulations and reality.

\subsection{The Three-point Correlation Function}

The highly non-Gaussian nature of the galaxy distribution can be
described with the hierarchy of $N$-point correlation functions. In
particular, the three-point correlation function contains a wealth of
information on gravitational collapse, initial conditions of the power
spectrum, and most importantly, bias and galaxy formation. In this
paper we present preliminary measurements of the three-point
correlation function in our DEEP2 GIF mock galaxy catalogs. Our goal
is to show that the three-point correlation function can be measured
with sufficient accuracy to be interesting and that the our selection
procedure does not systematically bias our results to a degree that
would hinder proper interpretation.

The three-point correlation function is defined as the reduced
joint probability of finding a triplet of galaxy at a particular
configuration \citep{pee80}. We have used the estimator by \citet{sza98}  
for extracting the three-point function:
\begin{equation} 
 \zeta = (DDD-3DDR+3DRR-RRR)/RRR.
\end{equation} 
In this equation contributions to a triplet from the data
and a random set are denoted with a $D$ and $R$ respectively.
The above estimator is expected to be the most accurate
edge-corrected formula in the family of estimators based
on triplets.

For our measurements of the GIF mock catalogs we chose eight
logarithmic bins centered on 0.613 $h^{-1}$ Mpc to 10.779 $h^{-1}$ Mpc.  All
possible triangles with those sides (allowed by the triangle
inequality) were used, i.e. 76 measurements were performed for each
sample. The fast $N$-point correlation code by \cite{moo00} based on a
novel double tree algorithm was used for the calculations.

We have divided the GIF mock galaxy catalogs into four redshift ranges 
$z$=0.7--0.9, 0.9--1.1, 1.1--1.3, and 1.3--1.5.
For each redshift range, we used the six independent catalogs to enable
a calculation of the cosmic (co)variances. Finally, for each simulation
we calculated the three-point correlation function for all the
galaxies in the catalog, as well as for only those galaxies which would
be selected to be observed using our target selection slitmask algorithm.
We used four times as many random points as galaxies for our
measurements.

\begin{figure} 
\epsscale{0.5}
\plotone{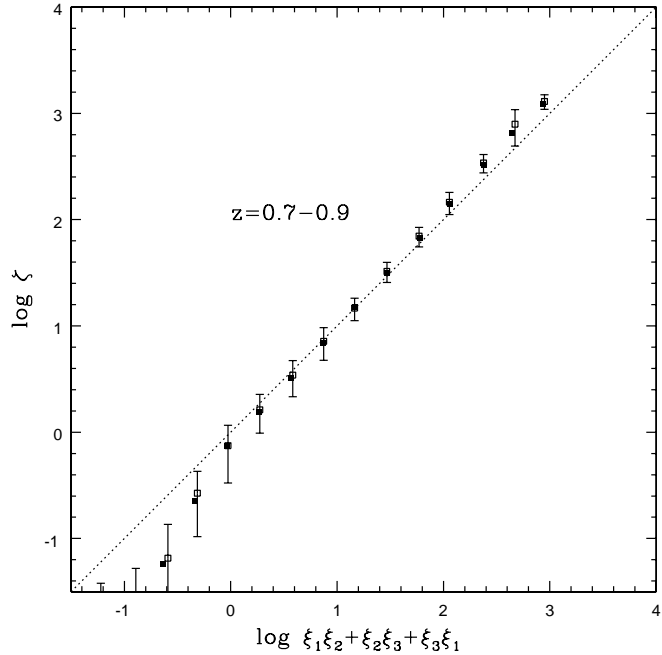}
\caption{The three-point correlation function as a function
of the hierarchical term  $\xi_1\xi_2+\xi_2\xi_3+\xi_3\xi_1$
in the redshift range $z$=0.7--0.9.
The open squares display the measurements for all the galaxies
in the mock catalogs, 
while the solid squares show measurements for only those 
galaxies targeted by the 2-pass slitmask algorithm. The data
were binned in terms of the heirarchical term, and the errorbars
correspond to the dispersion of the six mock catalogs as well
as scatter within the bin from one catalog.  The error bars on 
the solid squares are comparable to those displayed
 but are not shown for clarity.
The dotted line corresponds to a simple hierarchical
model with $Q_3=1$.} \label{3pt} 
\end{figure}

Figure \ref{3pt} contains the binned results for the shallowest slice,
$z$=0.7--0.9.  For this preliminary investigation the results 
are interpreted in terms of the simple hierarchical assumption 
\citep{dav83,pee80},
\begin{equation}
 \zeta = Q_3\left(\xi_1\xi_2+\xi_2\xi_3+\xi_3\xi_1\right).
\end{equation}
The three-point function is plotted in terms of the hierarchical
term on the right. If this equation were exact the result would be a 
straight line, as illustrated with a dotted line for the case $Q_3=1$.
Note that $Q_3$ shows a weak scale dependence.

According to figure \ref{3pt}, the hierarchical assumption gives
a good qualitative description of the data. The tight relation
illustrates the quality obtainable by the DEEP2 redshift survey.
The scatter is due to measurement errors, cosmic variance,
and an intrinsic shape dependence of the three-point correlation
function with scale.  The increasing error bars on large scales
is inevitable because of cosmic variance.  With a cross-scan size
of 20 $h^{-1}$ Mpc, triangles with baselines $>$10 $h^{-1}$ Mpc will
be poorly contrained by the DEEP2 survey.  Figure \ref{3pt} shows
that the effect of the target selection bias is to decrease $\zeta$
by $\sim$20\%.  This, of course, will have to be carefully modelled
in analyses of the real data.

Quantifying and separating the intrinsic shape dependence of the 
$\zeta$ with scale is the key to determining bias from the 
three-point correlation function.  Analysis on quasilinear scales
\citep{fry94} suggests that $\zeta$ can be used for a determination of the
galaxy bias that is independent from studies of kinematics or redshift-space
distortions.  In this way we will hopefully 
be able to measure absolute biases instead of relative biases.
This is left for a subsequent paper; here our purpose is mainly
to illustrate the feasibility of a high accuracy three-point measurement
with the DEEP2 redshift survey, and to show that the incompleteness 
caused by the slitmask algorithm does not cause significant, 
uncontrollable systematic errors.

\subsection{Measuring $\sigma_8$ and Relative Galaxy Bias 
as a Function of Redshift}

We expect to measure the evolution of $\sigma_8$ as a function of
redshift for the galaxies observed in the DEEP2 survey.  As a simple
test of the sensitivity of this measure, we use a
counts-in-cells analysis to calculate $\sigma_8$ in both our Hubble Volume
and GIF simulations mock galaxy catalogs in redshift bins of $\Delta z$=0.1.  
The counts-in-cells method \citep{pee80} uses the moments of the counts 
of numbers of galaxies in spheres of a given size to quantify the amount 
of galaxy clustering present.

By counting objects in randomly-placed spherical cells ($N_i$) of radius 
$l$, one can determine the moments of the counts:
\begin{equation}
\mu_1=<N_i>
\end{equation}
\begin{equation}
\mu_2=<(N_i-<N_i>)^2>
\end{equation}
For each mock catalog, we created 40,000 spherical cells with radii of 8 Mpc
$h^{-1}$ and randomly placed them throughout the catalog, ensuring that
 the centers of the cells were at least 8 $h^{-1}$ Mpc away from
the boundary of the mock catalog.  In this way none of the cells
overlapped the catalog boundaries.  We then divided the mock catalog
into slices of $\Delta z$=0.1 bins (eg. $z$=0.7--0.8) and calculated the 
moments of the galaxy counts and $\sigma_8$ as a function of redshift.
These moments are related to the volume-integrated two-point correlation
function and $\sigma_8$ as:
\begin{equation}
\mu_2=\mu_1+\mu_1^2\overline{\xi_2}
\end{equation}
\begin{equation}
\sigma_l=\sqrt{\overline{\xi_2}} .
\end{equation}

We performed this analysis on both the Hubble Volume and GIF mock catalogs,
as each has an emphasis on different properties which will be relevant 
in our DEEP2 survey data.  The GIF mock catalogs do not have continuous
evolution as they are made from three snapshot outputs at different redshifts
stacked in $z$-space.  However, we were able to use the absolute magnitudes
of the galaxies to apply a realistic magnitude-limited selection 
function, and as a consequence the galaxies at high-$z$ are intrinsically 
more luminous than those at low-$z$.  Since bias is expected to be 
luminosity-dependent, the {\it inherent} bias of the galaxies
 in the GIF catalogs will change with redshift.  The Hubble Volume 
catalogs, however, have realistic continuous evolution 
in their light-cone output, but the bias should change more slowly 
with redshift as the galaxy selection is the same at all redshifts.

To take into account the changing selection function with redshift, 
for the GIF mock catalogs we created a volume-limited 
sample within each redshift bin.  We used the absolute $B$-band magnitude to
keep only those galaxies which are bright enough so that they could 
have been observed at the far end of the redshift bin. For the Hubble 
Volume mock catalogs we did not
have magnitudes and so were not able to create volume-limited samples.
The Hubble Volume simulations therefore have a slight density gradient within 
each redshift bin.  We tested the effects of this small density gradient
on the counts-in-cells results from the GIF mock catalogs with and without
volume-limited samples and found a 1$\%$ difference in the values of
 $\sigma_8$ in the volume-limited sample, and therefore conclude that
it is a negligible effect.

\begin{figure}
\epsscale{0.5}
\plotone{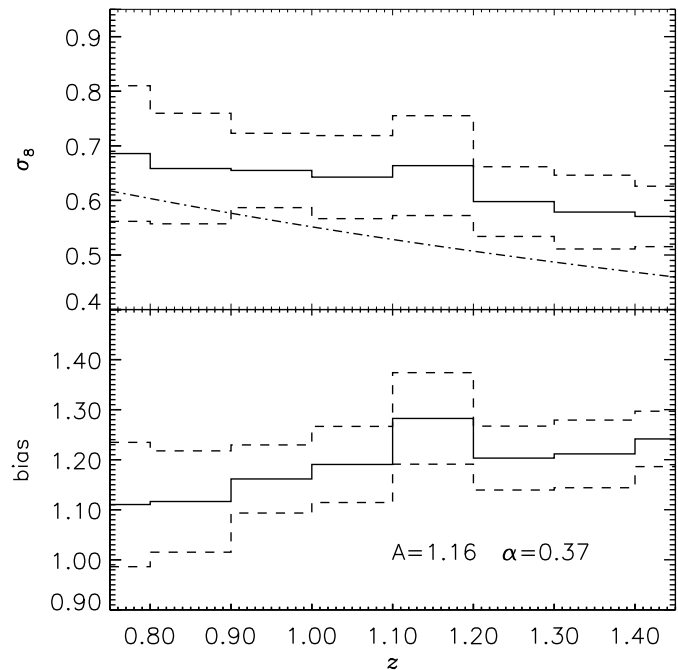}
\caption{a. The mean evolution of $\sigma_8$ as measured for galaxies
 in the Hubble Volume mock catalogs. 1$\sigma$ errors are plotted in 
dashed lines.  The dash-dot line shows the expected linear evolution of 
$\sigma_8$ of the underlying dark matter distribution, normalized to 
$\sigma_8$ of 0.9 at $z$=0.
b. The mean evolution of bias as defined by 
$\sigma_{8}$(galaxies)/$\sigma_{8}$(dark matter).  A fit
of bias=$A \left(\frac{1+z}{2}\right)^\alpha$ results in 
an amplitude $A$=1.16 and slope $\alpha$=0.37. \label{hubsigma8}}
\end{figure}

The results for the Hubble Volume mock galaxy catalogs 
are shown in figure \ref{hubsigma8}a.  The solid line is the mean
$\sigma_8$ as measured in our sixteen independent mock catalogs, while 
1$\sigma$ errors as determined from the variance between the different
catalogs are shown in dashed lines.  The expected
linear evolution of $\sigma_8$ in the underlying dark matter is shown
as a dash-dot line (normalized to 0.9 at $z$=0, as were the 
simulations; see Fry 1996 for analytic formula).  If we fit our estimates of $\sigma_8(z)$ of the galaxies to the
analytical form
\begin{equation}
\sigma_8(z)=A \left(\frac{1+z}{2}\right)^\alpha
\end{equation}
we find A=0.64 (1$\sigma$=0.01) and $\alpha$=--0.51 (0.10).  
The covariance between A and $\alpha$ is low;
the off-diagonal elements of the covariance matrix are equal to -0.36.

An evolution of the bias between the galaxies and the dark matter 
is apparent in the shallower slope of the galaxy
clustering evolution compared to the dark matter linear evolution 
prediction. Figure \ref{hubsigma8}b is a plot of the estimated bias,
equal to $\sigma_8$(galaxies)/$\sigma_8$(dark matter).  
The bias is slightly larger at high-$z$, as is expected from 
hierarchical structure formation models.  An analytic 
fit to the slope of the bias evolution results in a best fit of 
A=1.16 (1$\sigma$=0.01) and $\alpha$=0.37 (0.10), with the same low
covariance as the $\sigma_8$ analysis.  Evolution of bias
is therefore expected to be 
present at the 4$\sigma$ level within a single slice of the DEEP2 survey, but
separation of the changing bias from the evolving matter distribution will
be complex.  For a given cosmological model, we can use linear theory
to calculate the latter quantity.

\begin{figure}
\epsscale{0.5}
\plotone{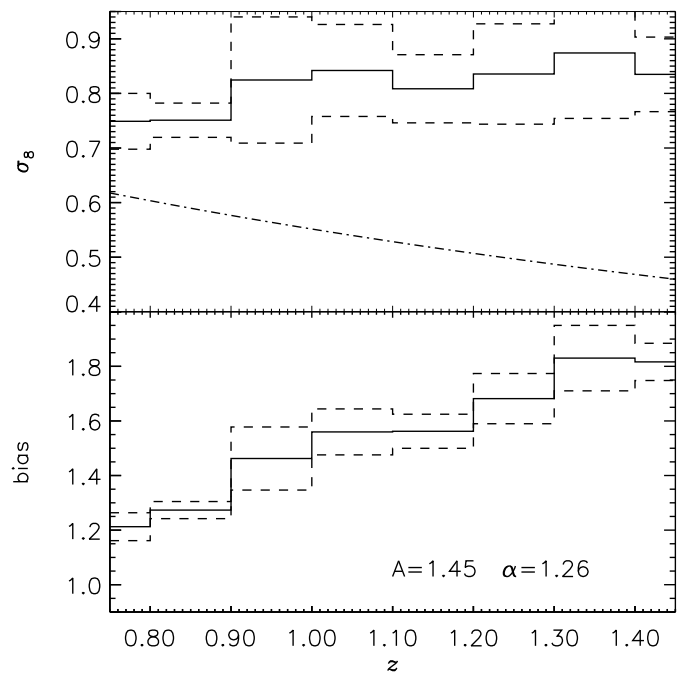}
\caption{a. The mean evolution of $\sigma_8$ as measured for galaxies
 in the GIF simulation mock catalogs. 
b. The change of bias with redshift.  Here the increase of bias
at high-$z$ is not solely the result of evolution, as 
 the galaxies at high-$z$ in our sample are more luminous and 
therefore have a higher degree of bias on the whole.
\label{gifsigma8}}
\end{figure}

The evolution of $\sigma_8$ for the GIF mock catalogs is shown in
figure \ref{gifsigma8}.  The best fit to the analytic form for the
evolution of $\sigma_8$ is A=0.80 (0.01) and $\alpha$=0.38 (0.13).
The change of bias with redshift is best fit by A=1.45 (0.01) and
$\alpha$=1.26 (0.13).
% these figures are made by sigma8_fit_nocov.pro
Here we are using mock galaxies of all luminosities, so that the
bias changes with redshift not primarily because of evolution but
because at high-$z$ we measure only the intrinsically more luminous galaxies 
in our magnitude-limited sample, and these objects are more clustered
at all redshifts.  This is an effect we will see in the
DEEP2 survey, where we will have to subdivide the galaxies by
linewidths or luminosity to separate the effects of bias evolution
from differing bias in different galaxy populations.  

\begin{figure}
\epsscale{0.7}
\plotone{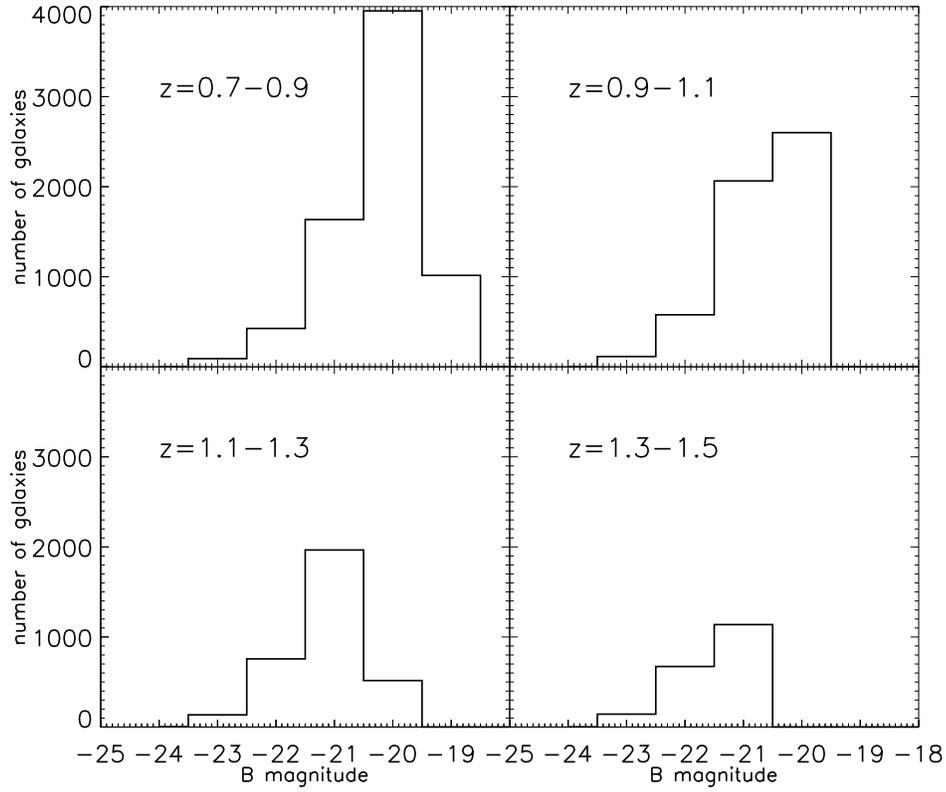}
\caption{The B-band luminosity function (LF) 
of one of our GIF mock catalogs in
different redshift bins.  The difference in the LF as
a function of redshift is due mainly to the selection effects of our 
apparent magnitude-limited sample, though there is some intrinsic 
evolution of the LF present in the simulations.  \label{lf}}
\end{figure}

\begin{figure}
\epsscale{0.7}
\plotone{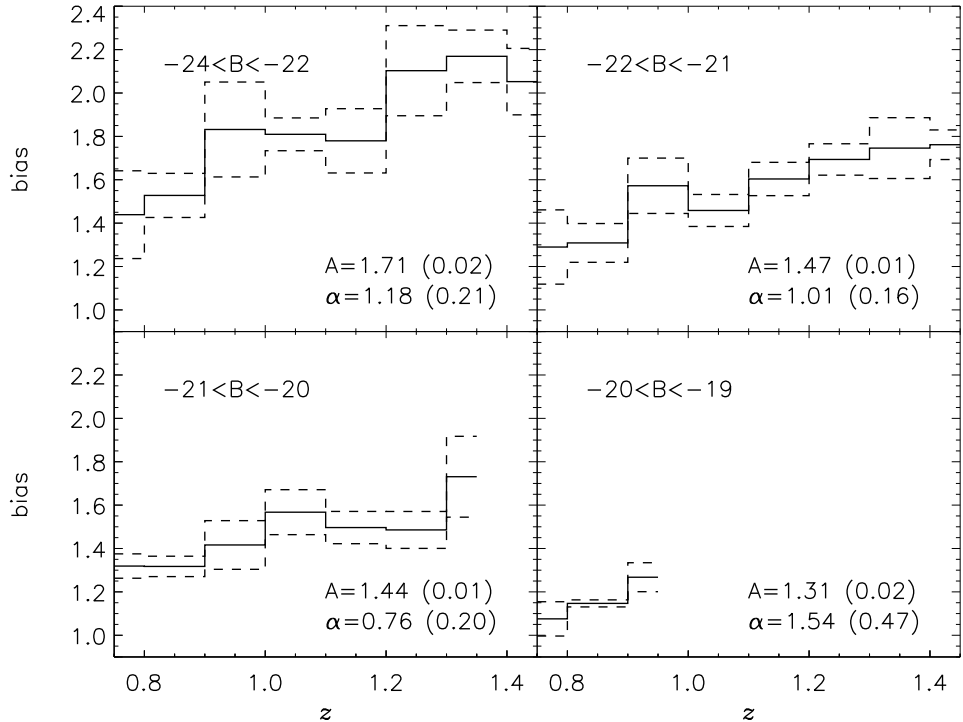}
\caption{Bias evolution for subsamples of the GIF mock catalogs selected
by B-band magnitude.   Here the change in bias with redshift is due
to evolution within a specific galaxy type. The brightest galaxies are
seen to have a larger bias and stronger evolution with $z$.  \label{bias}}
\end{figure}

In figure
\ref{lf} we show the B-band luminosity function (LF) of one GIF mock
catalog in different redshift bins.  At lower redshifts, the LF
extends to galaxies with absolute B mag of -19, but at higher redshifts
only the brightest galaxies are seen.  This change of the LF with
redshift is primarily due to our apparent magnitude sample selection,
though there is some inherent evolution in the LF which is included in
the GIF simulations.  We made subsamples of the galaxy catalogs based
on absolute B-band magnitude and show the bias evolution for
different magnitude ranges in figure \ref{bias}.  There is a clear
trend between galaxy magnitude and the amplitude as well as evolution of bias.
The brighter galaxies have a higher bias as well as stronger
evolution.  The values of the amplitude of the bias at $z$=1 (A)
and the slope of the evolution ($\alpha$) are given in the figure,
with 1$\sigma$ errors in parentheses.  We are able to measure
relative bias between galaxy types to within 0.02 and measure
evolution of bias at the 5-6$\sigma$ level for the brightest
galaxies (--24$<$B mag$<$--21) and at the 3-4$\sigma$ level for the fainter
galaxies (--21$<$B mag$<$--19).
Note that these values for the bias are calculated in redshift space.

\section{DISCUSSION}

A key goal of the DEEP2 redshift survey of galaxies is to constrain
the evolution of large-scale structure in the universe.  In most
cosmological models, the galaxy distribution is expected to evolve
measurably between $z=0$ and $z=1$, and different types of galaxies
are expected to evolve independently.  The survey will hopefully be
sufficiently robust to allow subdivision of the sample so as to
separately constrain the evolving bias of distinct classes of
galaxies. As indicated by the modest size of the error bars in the
comparison of mock catalogs, clustering constraints derived from the
DEEP2 survey are expected to have only modest cosmic variance.  They
will thus supersede the current contradictory constraints on the
strength of high redshift galaxy clustering derived from the modest
surveys undertaken to date.  As shown in figure 4, each slice of the DEEP2 
survey should lead to a 10\% estimate of the correlation length of 
clustering $r_0$.  Subdivision into differing samples will not degrade the
measurement precision as rapidly as $n^{-1/2}$, because cosmic   variance is
dominant in this estimate, not Poisson noise.  

The tests described above demonstrate the power of the DEEP2 survey for
refined estimates of redshift space distortions in the distant Universe.
  The high resolution of the DEIMOS spectra will yield high
quality redshifts, thus enabling the study of the kinematics internal to the
galaxies and the kinematics of small scale clustering at $z=1$.  It should
be readily possible to measure all these quantities within the survey volume,
although the estimated $\sigma_{12}$ is likely to be too noisy to set
useful constraints.   We note that $\sigma_1$ is measureable to 
10\% precision, while $\sigma_{12}$, after averaging over all 6 mock 
slices, is measurable to only 20\% precision.  The Kaiser infall effect 
on large scales should yield
a 10\% measurement of $\beta$, when averaging over the full survey.  Estimates
of the infall from subsamples
of the DEEP2 survey will not degrade as rapidly as the Poisson noise, 
since this statistic is also dominated by cosmic variance.

Precise constraints on the evolution of galaxy correlations will not only
greatly improve our understanding of galaxy formation in hierarchical
models, they will also facilitate novel cosmological tests of
considerable interest.  For example, the counting of galaxies in the
classic $dN/dz$ test (Newman \& Davis, 2000; 2001) is capable of
setting good constraints on cosmological parameters such as $w\equiv
P/\rho$ versus $\Omega_m$, but serious degeneracy will remain even
after counting $10^4$ galaxies in the range $0.7 < z < 1.5$.  The
magnitude-redshift relation derived from distant supernovae leads to
constraints with very similar degeneracy between $w$ and $\Omega_m$.
To a considerable degree this degeneracy can be broken by additional,
complementary tests.  For example, the expected ratio of the $J_3$
integral of $\xi(r)$ at $z=1$ compared to that at $z=0$ has a different
dependency on these same cosmological parameters, leading to contours
of constant $J_3$ ratio that cross those of the $dN/dz$ test.  Further
tests, such as the abundance of clusters of galaxies, lead to quite
different constraints in this same parameter space plane (Newman et al.
2001).  Simultaneous execution of all these tests with the DEEP2
survey data is thus capable of setting rather tight constraints on $w$
and $\Omega_m$.

\section{CONCLUSIONS}

The clustering of galaxies at high redshift
is intimately connected to our incomplete 
understanding of galaxy formation, and both disciplines will advance 
 as we analyze the details of clustering  in the soon-to-begin DEEP2 survey.
 In order to study the precision with which one can
expect to measure the clustering of galaxies at $z \sim 1$, 
we have generated a series of mock catalogs of the DEEP2 galaxy
redshift survey.   The mock
catalogs are derived from two separate VIRGO simulations, one with
continuous evolution of the underlying structure but inadequate mass
resolution, and another with adequate mass resolution but which
required the stacking of numerous simulation volumes to achieve a size
sufficient to match the redshift range of the survey.

The DEEP2 survey will use multislits, rather than multifibers, to
multiplex spectroscopic targets.  The necessity of avoiding
overlapping spectra in the survey limits the sampling efficiency of
spectroscopic target selection, especially in regions of the sky with
higher than average surface density of galaxies.  The tentative
selection algorithm for the survey will lead to $\sim$ 70$\%$ success
rate in target selection.  Although the resulting bias substantially
affects the angular correlations of the selected targets, the
correlations observed in redshift space are only modestly affected,
 and the bias can be largely
removed by double counting galaxies closest on the sky and within the 
photometric redshift error to those
galaxies not selected for spectroscopy.

Distortions in redshift space as measured by the $\xi(r_p,\pi)$
diagram will be cleanly detected in the DEEP2 survey, and we expect to
measure both the "fingers of god" on small scales and coherent infall
motions of galaxies on large scales.  Additional tests of velocity
field effects, such as the $\sigma_1$ test of \citet{dav97}, will
be readily possible with the sample and will be much less affected by
the bias in the target selection algorithm. 

We expect to be able to measure  $\sigma_8$ to within 10\% in individual slices
of the survey.  Assuming a cosmological model, we can 
detect the evolution of relative bias with 
4--5$\sigma$ significance within the volume of the DEEP2 survey
itself, with more precise measurements for the brightest subsample
of galaxies.  

The survey geometry is designed to allow a good measure of the three-point 
clustering amplitude for scales $r \le 10h^{-1}$ Mpc, and tests
of the mock catalogs indicate that this will indeed be possible.  Such
a measure will allow an independent test of the bias in the galaxy
distribution.  

The DEEP2 survey is designed to be a comprehensive study of the Universe
at $z=1$.  Having a sample of high redshift galaxies comparable in quality to 
those available locally will enable many independent tests of the
evolution of structure in the Universe.  With numerous cross checks and
subdivision of the redshift catalogs into different classes of galaxies, we
should be able to untangle evolution in the galaxy bias from evolution in the
underlying matter distribution.

\

We would like to thank Guinevere Kauffmann for graciously allowing us to use
her semi-analytic GIF simulations.  IS would like to thank his 
collaborators on {\tt npt} (the fast $N$-point code), especially 
Andrew Connolly, Bob Nichol, Andrew Moore, and Alex Szalay, for their 
help. We acknowledge helpful discussions with Jeff Newman.
This project was supported by the NSF grant AST-0071048.  
A.L.C. would like to acknowledge an NSF Graduate 
Research Fellowship.  IS was partially supported by a NASA AISR, NAG-10750.
 We also acknowledge the support of Sun Microsystems.

\end{document}